\documentclass[12pt, preprint]{aastex}
\begin{document}

\bibliographystyle{apj}

\def\Msun{M$\odot$}

\title{Dust around Type Ia supernovae}
\author{Lifan Wang$^{1,2}$}
\affil{$^1$Lawrence Berkeley National Laboratory, 
1 Cyclotron Rd, Berkeley, CA 94712\\
$^2$ Purple Mountain Observatory, Nanjing, China
}

\begin{abstract} 
An explanation is given of the low value of 
$R_\lambda \equiv A_\lambda/E(B-V) $, the ratio of absolute to 
selective extinction deduced from Type Ia supernova observations. 
The idea involves scattering by dust clouds located in 
the circumstellar environment, or at
the highest velocity shells of the supernova ejecta. 
The scattered light tends to reduce the effective $R_\lambda$ in
the optical, but has an opposite effect in the ultraviolet. 
The presence of circumstellar dust can be tested by ultraviolet 
to near infrared observations and by multi-epoch spectropolarimetry 
of SNe~Ia. 
\end{abstract}

\keywords{Cosmology: distance scale --- dust --- supernovae: general}

\section{Introduction}

  Dust extinction of SNe~Ia is of critical 
importance to supernova cosmology. New studies show that 
when correcting the observed $B$ magnitudes of SNe~Ia to 
the observed $B-V$ colors, the coefficient is generally 
found to be around 2-3 instead of the 
value of about $R_B \equiv A_B/E(B-V)=4.1$ as 
expected for dust extinction in the Galaxy and the LMC  
\citep{Tripp:1999, Phillips:1999, Wang:2003, Knop:2003, Wang:2005}. 
This small value had been thought to be likely caused by a combination of intrinsic 
color dependence of SN luminosity and dust extinction. 
Some recent works however show that for several well 
observed highly extinct SNe~Ia, $R_B$ were found to 
be in the same low range as deduced from cosmology fits 
({\it e.g.} \cite{Krisciunas:2001, Krisciunas:2004}). 
These studies seem to suggest that dust in SN hosts
are systematically different from those in the Galaxy 
and the LMC. Here I propose an alternative explanation 
that does not require such a difference. 

\section{The circumstellar dust of SNe Ia}

There might be circumstellar (CS) dust around the progenitor 
systems of SNe~Ia. In such cases extinction correction can not
be performed by assuming a standard interstellar extinction law,
but has to be treated through careful radiative transfer \citep{Witt:1992}.
SN~2002ic, as an extreme example, 
was found to be associated with a massive hydrogen rich material 
with mass around 6$ {\rm M\odot} /(n/10^8 cm^3)$
\citep{Hamuy:2003, Wang:2004, Wood-Vasey:2004, Deng:2004}. 
\cite{Wang:2004} deduced from spectropolarimetry 
observations that the hydrogen rich materials are
distributed in an asymmetric, perhaps disk-like geometry. 
Such a massive envelope, if exists in the Galaxy, must 
be easily observable. It can in fact be identified with 
well studied post-AGB objects such as proto-planetary nebulae (PPNe). 
The post-AGB phase is very short lived, lasting only 
on the order of a few thousand years. This explains why 
SN~2002ic like events are 
rare but it also raises the question of whether SNe~Ia can occur
in environment in which the surrounding nebulosity is more diluted.
PPNe ultimately evolve to planetary nebulae (PNe) which then 
disperse into the interstellar medium (ISM) and leave behind 
white dwarfs in the center. A large fraction of white dwarfs 
inside PNe are found to be likely in binary systems 
\citep{DeMarco:2004}. The PN phase lasts for about ten to 
a hundred times longer than the PPN phase. Assuming that 
the explosion of the central white dwarfs are unrelated to 
the evolution of nebulosity outside, one can expect that 
there are about 10 times more SNe~Ia occurring inside PNe for 
every SN~Ia occurring inside a PPN.  Extinction to the central 
white dwarfs of several PNe were observed by 
\cite{Wolff:2000}, and the dust extinction optical 
depth is typically around 1. Assuming that there is 
no dust creation/destruction after the PPN 
shell ejection and homologous expansion of the 
nebulae, at any later epochs the dust opacity scales 
as $ \frac{\tau_B}{1} (\frac{10000 {\rm year}}{t})^2$, where $t$ is 
the dynamical age of PNe which is typically around 
10,000 years.  It thus takes about 
$ 10^5$ years for the dust to be diluted to $\tau < 0.01$ - 
a level that is still sensitive to modern SNe Ia observations. 
Dusts in PNe are distributed in patchy opaque clumps 
such as observed in the Helix nebula \citep{ODell:2004}. These 
dusty clumps may survive even longer time scales.
On top of the dust ejected as PPNe during the post-AGB phase, 
more dust may be ejected to the CS environment throughout
the evolution path to SNe~Ia. This argues that circumstellar 
dust may be important and has to be analyzed carefully for precision 
measurements.

\section{Dust scattering and absorptions}

The albedo of interstellar dust is around 0.7 
in $B$ and $V$ filters, as was found 
from observations of reflection nebulae (see \cite{Draine:2003} 
for a review of interstellar dust properties). This means that 
scattering dominates the interaction between photons and dust particles.

\subsection{The case of  time invariable sources}
For illustrative purposes, let us consider an invariant point 
source located inside a optically thin CS shell of 
optical depth $\tau \ll 1$ and albedo $\omega$. In one 
extreme case (Case A, hereafter) we assume that all the scattered 
photons do not reach the observer, then the attenuation of the source
is given by $e^{-\tau}$. Case A applies to interstellar dust 
extinction where scattering predominantly direct photons off 
the line of sights. If on the other extreme (Case B, hereafter) 
we assume that all of the scattered photons eventually 
escape from the system, the corresponding attenuation 
will be $e^{-\tau \cdot (1-\omega)}$, where it is also assumed
that each photon interacts at most once with the 
dust shell before escape which is a good approximation 
if the shell is optically thin. Case B alters the extinction 
curves of the dust from 
$A_\lambda$ to $A^0_\lambda = A_\lambda(1-\omega(\lambda))$. 
A complete description of the amount of extinction requires not only 
the extinction cross section but also the dust albedo. Using the 
interstellar dust model of \cite{Weingartner:2001} for the average 
properties of LMC dust, the two limiting cases are shown 
in Fig. 1. The inclusion of scattered 
photons not only reduces the total extinction
but also changes significantly $R_\lambda \equiv\ A_\lambda/E(B-V)$, 
the ratio of extinction to color-excess. $R_\lambda$ 
is significantly reduced at wavelength longer than 300 nm 
whereas the opposite is true in the wavelength range shorter than 300 nm. 

\begin{figure}
\figurenum{1}
\epsscale{0.7}
\plotone{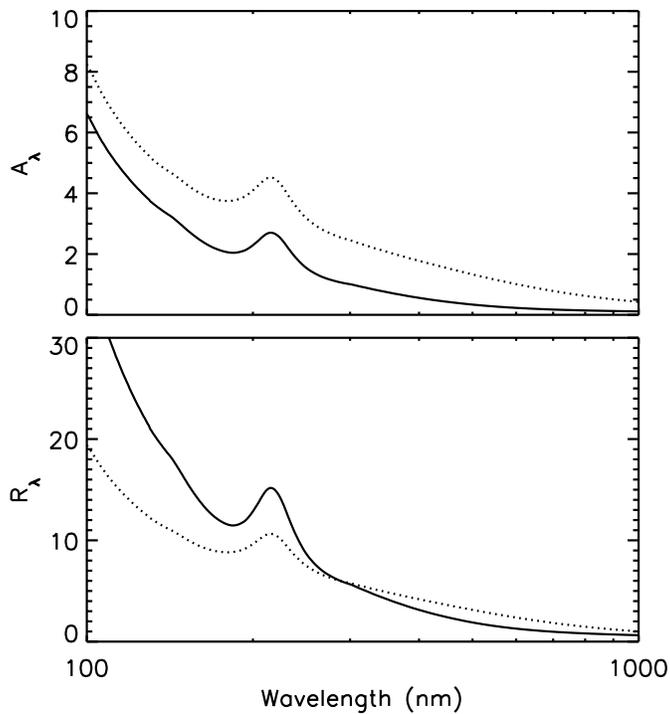}
\caption{The extinction properties of the dust model of 
\cite{Weingartner:2001} for the LMC average. The dotted lines show the
the total extinction (a, upper panel) and the ratio of total extinction
to E(B-V) (b, lower panel) if scattered photons are all 
unobservable (Case A, see text).
The solid lines show the corresponding quantities assuming single 
photo-dust interaction and all scattered photons escape from the 
dust shell and are observable (Case B, see text). The 
extinction curves are dramatically different for Case A and Case B.
}
\end{figure}

Case A is applicable for extinction by interstellar dust. Case B 
is applicable for extinction by a dust envelopes that can not 
be spatially resolved from the target, such as compact CS 
dust of stars or circumnuclear dust 
in the host galaxies of AGNs or QSOs. 

\subsection{The dependence of dust attenuation on the spectral evolution of SN~Ia} 

The light curves and spectra of SNe evolve with time. The 
effective wavelengths in different filters thus 
vary, and accordingly the amount of 
extinctions in these filters change with time. 
Using the dust model of \cite{Weingartner:2001}, 
and the SN~Ia spectral template as described in \cite{Knop:2003}, we 
show in Fig.~2(a)  $R_\lambda$ at different epochs 
for a typical SN~Ia. R$_\lambda$ show $\sim$ 20\%\ variations. 
This effect should be important when 
using SN~Ia for cosmology, but can not explain the observed 
low $R_B$ for SNe~Ia.

 \begin{figure}
\figurenum{2}
\epsscale{0.7}
\plotone{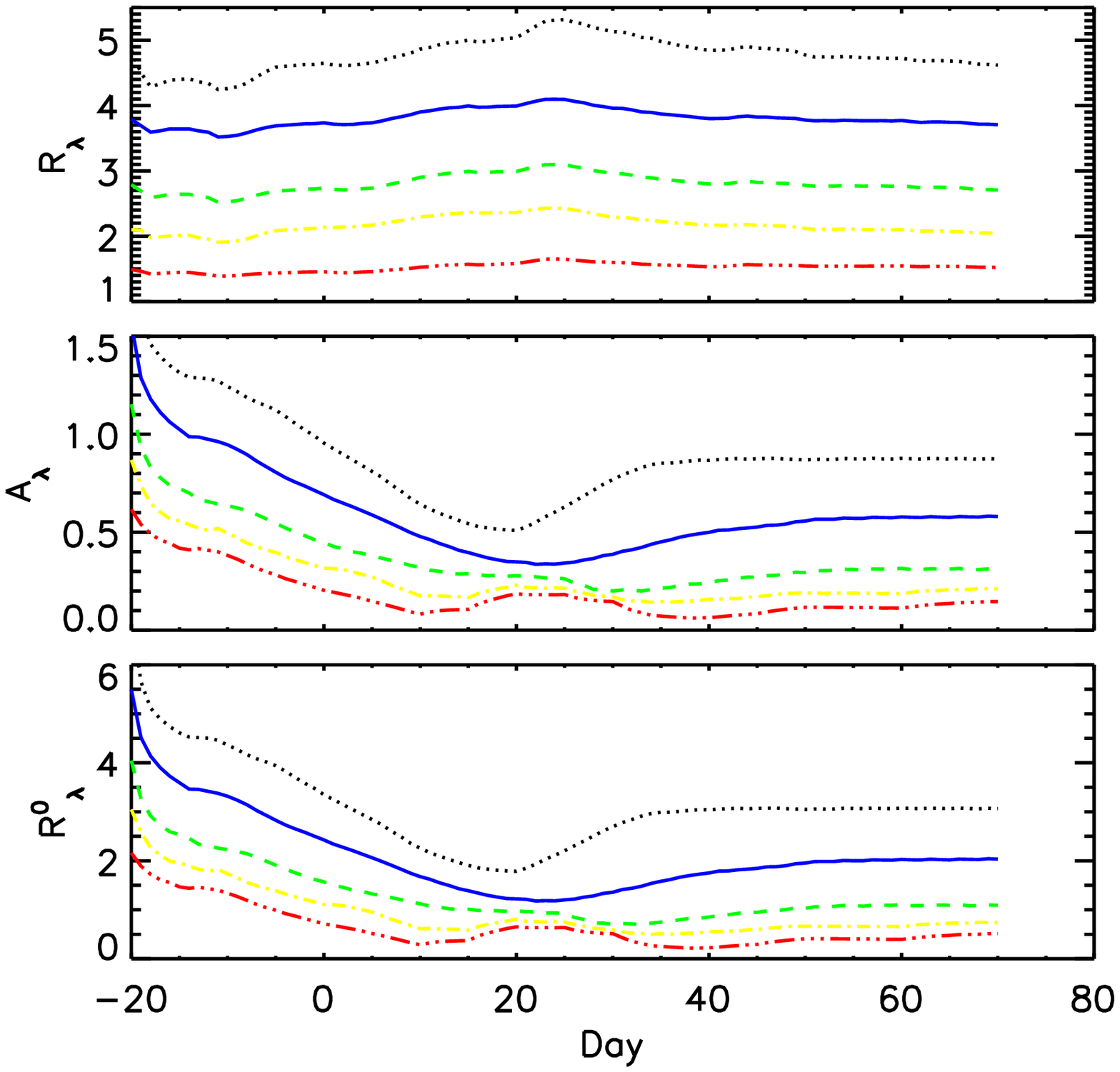}
\caption{The time dependence of 
extinction properties in $U$ (dotted line), 
$B$ (solid line), $V$ (dashed line), $R$ (dot-dashed line), 
and $I$ (dot-dot-dot-dashed line) bands for Type Ia supernovae. 
(a), upper panel, shows the effect on $R_\lambda$ 
due to spectral time evolution. (b), middle panel, and (c), lower panel, 
show the $A_\lambda$ and $R^0_\lambda$ as defined in \S3.3, respectively.
The curves in (b) and (c) were derived assuming a spherically symmetric
distribution of dust at a radius above 1 $10^{16}$ cm and 
an optical depth of 1.45 in $B$ band.  
}
\end{figure}

\subsection{The time dependency of dust scattering}
Light reflected off dust particles travels longer distance and 
arrives at the observer with a time delay. This is the 
so called light echo phenomena studied by \cite{Chevalier:1986}, 
and more recently by \cite{Sugerman:2003} and \cite{Patat:2005}. 
Light echos are observed in several nearby SNe such as 
SN~1987A \citep{Crotts:1989, Sugerman:2005}, 
SN~1991T \citep{Schmidt:1994}, and SN~1998bu \citep{Cappellaro:2001}. 
The survival of CS dust around SN~1987A was studied by 
\cite{Wang:1996} where it was shown that scattering by 
circumstellar dust can provide an alternative explanation of
early polarimetry of SN~1987A. \cite{Wang:1996b} also suggest that
dust around SNe~Ia can be probed by polarimetry observations.

The exact amount of extinction is related to the geometric 
location of the dust clouds. In Fig. 2(b) and (c) we 
show as an example the results assuming a geometry in which 
dust coexists with a stellar wind of inner radius 1 10$^{16}$ cm. 
The optical depth of the dust cloud is assumed to be 1.45 in $B$ 
band which gives $A_B = 1.34\ mag$. 
The effect of multiple scattering is treated approximately 
using the formula of \cite{Mathis:1972}. With the 
extinction cross section as given by \cite{Weingartner:2001} 
for dust in the LMC, the column density of the dust shell 
considered here is 10$^{22}$ cm$^{-2}$ which requires 
a mass loss rate of $3.3\cdot 10^{-5} {\rm M\odot/year}$.

As shown in Fig. 2(b), the extinctions in different filters 
decrease steadily from the time of explosion to about 15-20 days 
after optical maximum. This is due to an increase of the 
contributions of scattered photons to the total flux. The 
effective extinction is in general smaller than when dust 
scattering is ignored. Fig. 2(c) shows the effective ratio 
of extinction to color-excess. To be consistent with SN~Ia 
observations, this is defined here as 
$ R^0_\lambda \ = \ A_\lambda/E^0(B-V)$, with 
$E^0(B-V)\equiv(A^{max}_B-A^{max}_V)$, 
where $A^{max}_B = B^{max}_d - B^{max}$ and 
$A^{max}_V = V^{max}_d-V^{max}$ with $B^{max}$ and $V^{max}$ being the 
$B$ and $V$ band maximum magnitudes with no dust extinction, 
respectively, and $ B^{max}_d$ and $V^{max}_d$ the $B$ and $V$ 
band maximum magnitudes with dust extinction, respectively. 
By comparing Fig. 2(a) and Fig. 2(c), it is remarkable 
that the inclusion of dust scattering reduces significantly 
the values of $R_\lambda$ around optical maximum in the 
optical wavelength range.

\begin{figure}
\figurenum{3}
\epsscale{0.7}
\plotone{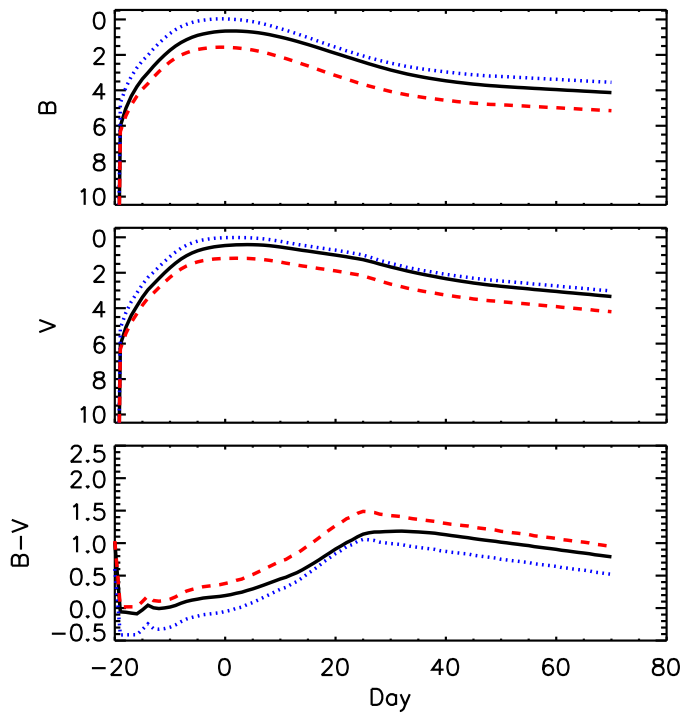}
\caption{The $B$ (upper panel), $V$ (middle panel) light curves 
and $B-V$ color curves (lower panel) of Type Ia supernova. The dotted
lines show the input data. The dashed lines show the light curve expected
for Case A where scattered photons are all unobserved. The solid lines
show the results when scattering and absorption are treated properly by
inclusion of the time delayed scattered photons. The geometry 
and opacity of the dust are the same as in Fig. 2.}
\end{figure}

The $B$ and $V$ band light curves and the color 
curve $B-V$ are shown in Fig. 3. The presence of 
dust alters the light curves shapes, results in light curves 
with steeper rises and flatter decreases. This affects 
the measurements of light curve parameters. It is worth
pointing out that the peculiar SN~2000cx showed 
qualitatively a fast rise and a slow decline in
$B$ and $V$ bands \citep{Li:2001, Candia:2003} which 
is consistent with the above behavior. We defer 
quantitative analyses of individual SN to future studies, but
note here that the presence of CS dust can be 
tested on individual basis of well observed SNe. 

Fig. 4 shows that scattering has smaller effect around 
optical maximum for dust wind located at larger distances.
More distant dust clouds, however, do affect 
later time light curves.

\begin{figure}
\figurenum{4}
\epsscale{0.7}
\plotone{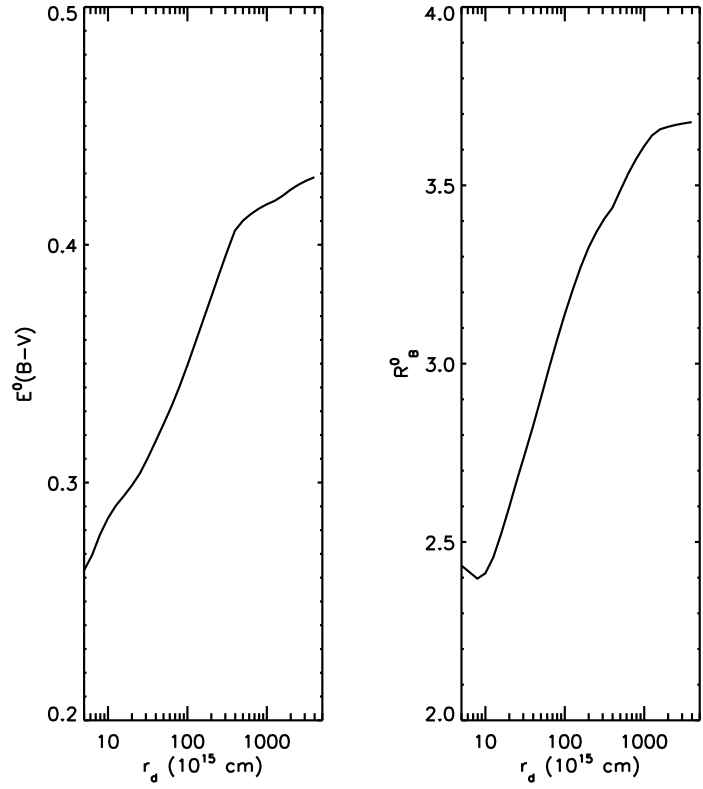}
\caption{The effect of dust scattering as a function of the inner boundary
of the dust clouds. (a), left, shows the color-excess, and (b), right, shows
$R^0_\lambda$ as defined in \S 3.2.}
\end{figure}

\section{Discussions and conclusions}
 This study shows that the low values of $R_B$ observed SNe~Ia  
may be an indication of the presence of dust in the immediate 
neighborhood of SNe~Ia. Dust of extinction properties similar to 
that of the LMC can explain the observed unusually low  
$R_B$ values for SNe~Ia, provided that they are distributed in 
the circumstellar environment of the SNe. Although not required 
by the current analyses, it should be reminded that CS dust may 
have substantially different extinction properties compared to 
interstellar dust. However, the presence of even a small amount 
of CS dust leaves clear imprints on the light curves of SNe~Ia 
and can be tested by careful observations of SN~Ia light curves.

High resolution spectroscopy has set some upper limits 
on the amount of CSM around SN~1994D \citep{Cumming:1996} and SN~2001el 
\citep{Mattila:2005}. The strongest constraint is a mass loss rate of 
around 10$^{-5} {\rm M\odot/year}$ for a wind velocity of 10 km/s 
derived for SN~2001el 9 days before optical maximum. 
Note that at this early date, only CSM at distances lower 
than 3 $10^{15}$ cm are interacting with the SN ejecta, 
and the observations are not sensitive to CSM at even 
larger distances. These observations are thus 
insensitive to nebular structures with a central bubble 
such as often encountered in PNe. 
 
As noticed in \cite{Wang:2004} and \cite{Deng:2004}, several other 
SNe previously identified as SN~IIn are in fact strikingly 
similar to SN~2002ic at late stages. These include 
SN~1988Z \citep{Turrato:1993}, SN~1997cy \citep{Turrato:2000}, 
and SN~1999E \citep{Rigon:2003}. Model spectra of these objects
seem to rule out significant amount of oxygen in the ejecta 
\citep{Chugai:1994, Turrato:2000, Chugai:2004}.
\cite{Chugai:1994} suggested that the mass of SN~1988Z ejecta
is unexpectedly low with M $<\ 1 {\rm M\odot}$. The low mass ejecta
and the absence of oxygen are consistent with SN~Ia explosions. 
If these are all 
SN~2002ic-like SNe, and taking the number of SN~Ia discoveries
at their face value, it would imply that about 1\%\ of all SNe~Ia
are associated with dense nebula similar to SN~2002ic. 
There may be a substantial fraction of 
SNe~Ia with significant amount of undetected 
CS materials. This is corroborated by recent SN~Ia rate studies 
which indicate that about 50\%\ of the observed SNe~Ia are produced by
progenitors probably more massive than 5.5 M$\odot$, 
in a time scale of the order of 10$^{8}$ years after the 
progenitor birth \citep{Mannucci:2005}.
\cite{Livio:2003} argue merging of two white dwarfs might be 
responsible for events such as SN~2002ic, whereas \cite{Chugai:2004}
argue for the Type 1.5 SN scenario proposed by \cite{Iben:1983}
in which the explosion is due to a star at the end of the
post-AGB phase \citep{Iben:1983}. These models would produce SNe~Ia 
with dense CSM envelope. It is not clear in what parameter range
can these different models produce successful supernovae. Studies 
of CS dust around SNe~Ia can be performed and be used as probes 
of the progenitor systems and explosion mechanisms. 
   
  Another source of dust in the immediate 
neighborhood of SNe~Ia may be the ejecta themselves. The thermonuclear 
reaction lasts only for a second after the SN explosion. During which 
the ejecta reach a temperature of 10$^9$ K and an expansion velocity 
of 25,000 - 30,000 km/sec. The temperature of the ejecta decreases 
rapidly due to adiabatic cooling and reaches a temperature of about 
$10^3$ K in only about a few minutes. This rapid cooling 
allows for the condensation of dust in the ejecta.  Most of 
these dust, however, are likely to be quickly destroyed by 
radioactive heating as the ejecta are reheated to temperatures 
around 10,000 K.  But the dust formation may be  patchy 
and some dense clumps may survive the radiation field of 
the supernova, especially at the highest 
velocity layers that are shielded from the bombardment 
of the radioactive decays deeper. 

In summary, the signatures of CS dust can be 
recognized by analyzing light curves of SNe~Ia. Observations of 
wide wavelength coverage from UV to the near-IR  offer 
the best hope for discriminating interstellar dust from CS dust. 
Spectropolarimetry is another method for studying interstellar
dust, as shown in \cite{Wang:1996}
for SN~1987A.

{\bf Acknowledgment}  I am grateful to 
G. Aldering, S. Perlmutter, M. Strovink, A. Witt, and D. York for helpful 
discussions.


\end{document}